\documentclass[structabstract]{aa} 
\usepackage{tabularx}                                  
\usepackage{enumerate}
\usepackage{graphicx}
\usepackage{txfonts}
\usepackage{multirow}
\usepackage[sort]{natbib}

\begin{document}

   \title{Abell 2384: the galaxy population of a cluster post-merger\thanks{Tab. 4 is only available in electronic form at the CDS via anonymous ftp to cdsarc.u-strasbg.fr (130.79.128.5) or via http://cdsweb.u-strasbg.fr/cgi-bin/qcat?J/A+A/}}

   \author{Florian Pranger\inst{1}\fnmsep\thanks{\email{florian.pranger@uibk.ac.at}}
          \and Asmus B\"{o}hm\inst{1}
          \and Chiara Ferrari\inst{2}
          \and Sophie Maurogordato\inst{2}
          \and Christophe Benoist\inst{2}
          \and Harald H\"{o}ller\inst{1}
          \and Sabine Schindler\inst{1}
          }

   \institute{
Institute for Astro- and Particle Physics, University of Innsbruck, Technikerstr. 25/8, A-6020 Innsbruck, Austria
\and Laboratoire Lagrange, UMR7293, Universit\'{e} de Nice Sophia Antipolis, CNRS, Observatoire de la C\^{o}te d'Azur, 06300, Nice, France
}          
   \date{Received \today; accepted ???}
 
  \abstract
  {We present a spectrophotometric analysis of the galaxy population in the area of the merging cluster Abell 2384 at z=0.094.}   
   {We investigate the impact of the complex cluster environment on galaxy properties such as colour, morphology and star formation rate.}
   {We combine multi-object spectroscopy from the 2dF and EFOSC2 spectrographs with optical imaging of the inner 30x30 arcminutes of A2384 taken with the ESO Wide Field Imager. We carry out a kinematical analysis using the EMMIX algorithm and biweight statistics. We address the possible presence of cluster substructures with the Dressler-Shectman test. Cluster galaxies are investigated with respect to [OII] and H$\alpha$ equivalent width. Galaxies covered by our optical imaging observations are additionally analysed in terms of colour, star formation rate and morphological descriptors such as Gini coefficient and M$_{20}$ index. We study cluster galaxy properties as a function of clustercentric distance and investigate the distribution of various galaxy types in colour-magnitude and physical space.}
   {The Dressler-Shectman test reveals a substructure in the east of the 2dF field-of-view. We determine the mass ratio between the northern and southern subcluster to be $\lesssim$1.6:1. In accordance with other cluster studies, we find that a large fraction of the disk galaxies close to the cluster core show no detectable star formation (SF). Probably these are systems which are quenched due to ram-pressure stripping. The sample of quenched disks populates the transition area between the blue cloud and the red sequence in colour-magnitude space. We also find a population of morphologically distorted galaxies in the central cluster region.}
   {The substructure in the east of A2384 might be a group of galaxies falling onto the main cluster. We speculate that our sample of quenched spirals represents an intermediate phase in the ram-pressure driven transformation of infalling field spirals into cluster S0s. This is motivated by their position in colour-magnitude space. The occurrence of morphologically distorted galaxies in the cluster core complies with the hypothesis of Abell 2384 representing a post merger system.}

               
\authorrunning{F. Pranger et al.} 
\titlerunning{Abell 2384: the galaxy population of a merging cluster system}

\maketitle
%

\section{Introduction}
In the local universe galaxies are found to populate environments of different density, going from voids, galaxy groups and filaments, to the inner regions of galaxy clusters (e.g. \citealt{boselli06}). Ever since the fundamental work by \citet{dressler80}, connections between a variety of galaxy properties and the density of their surrounding environment have been proposed and analysed (e.g. \citealt{bamford09}). By studying the transition region between the cluster outskirts and the inner and denser regions, it is possible to directly compare galaxies in environments of different density. Moreover, since massive galaxy clusters have not finished their growing phase up to the present epoch (e.g. \citealt{donnelly01}), merging clusters offer the possibility of studying complex dynamics and the influence of merger related processes on their galaxy populations. It is, for example, still a matter of discussion, whether (and, if so, how) cluster mergers influence star formation activity within cluster members. While e.g. \citet{shim11} deduce from their MIR observational results that the merging process suppresses star formation in the galaxies of the merging cluster Abell 2255 (z$\sim$0.08), \citet{bekki10} find in numerical SPH simulations of cluster mergers that a merger event can trigger star formation episodes in gas rich galaxy halos due to a significant compression of their cold gas by the increased external (i.e. ICM) pressure. In a recent study on Abell 2744 \citet{rawle14} find a population of star forming galaxies which show a "jellyfish" morphology, likely due to the passage of a merger shock front. Comparing pre- and post-shock galaxies in the "Bullet cluster", \citet{chung09} find possible hints (at $\sim$2$\sigma$ significance level) of reduced specific star formation rates in post-shock galaxies.  

Interactions between two or more galaxies due to close spatial encounters can lead to morphological distortions such as tidal tails and bridges, bars or warps. Such galaxy-galaxy interactions are most efficient at low relative velocities as for example in groups or pairs of galaxies or in the outer regions of galaxy clusters (e.g. \citealt{toomre72}). However, as shown by e.g. \citet{kleiner14} cluster mergers can lead to a more frequent occurrence of galaxy-galaxy interactions and galaxy mergers in the inner regions of clusters which can trigger star formation episodes in the galaxies (e.g. \citealt{ferrari05}).\\ Another process potentially affecting the morphology, gas content and gas distribution of a galaxy in the inner cluster regions is galaxy harassment \citep{moore96}. This term applies to galaxies being exposed to tidal forces due to the cluster's gravitational potential and consecutive fly-bys with other cluster members.\\
Simulations have confirmed that violent galaxy-galaxy interactions can trigger episodes of star formation in the involved systems (e.g. \citealt{mihos96}).\\

Apart from dynamical galaxy-galaxy interaction processes hydrodynamic interactions between the gaseous content of a galaxy and the hot gas trapped in the cluster's potential (i.e. intra cluster medium or ICM) have been revealed and investigated. A galaxy moving relative to the ICM experiences ram-pressure which can eventually remove (or strip) the galaxy's gaseous halo and disk (see e.g. \citealt{abadi99}). Simulations by e.g. \citet{kapferer09} and \citet{steinhauser12} have shown that ram-pressure stripping can enhance star formation for a short time period, while e.g. \citet{quilis00} demonstrate that on long time scales ram-pressure stripping leads to quenching of star formation.\\  

In the course of their investigations of galaxy clusters a number of authors describe populations of non star forming disk galaxies (e.g. \citealt{poggianti99, goto03, koopmann04, vogt04, pranger13}). As part of their analyses of a volume-limited sample of SDSS data \citet{goto03} find that such galaxies mainly occur in high-density environments, i.e. near the centre of galaxy clusters. Regarding the high probability of dynamical interaction processes (e.g. galaxy mergers or galaxy-galaxy tidal interactions) to affect the morphology of the stellar disk the authors propose a cluster related mechanism to explain the stellar disks of the non star forming galaxies being found undisturbed. \citet{vogt04} argue that the non star forming disk galaxies could represent an intermediate stage of a ram-pressure driven morphological transformation of spiral galaxies (falling into the cluster from the field) into cluster S0s. Another intermediate phase in this transition that occurs before the final cessation of star formation might be defined by the class of red spiral galaxies as described by \citet{wolf03}. These galaxies show average specific star formation rates (SFRs) four times lower than blue spirals \citep{wolf09} and are interpreted as the low specific SFR tail of the blue cloud \citep{boesch13}. They populate the so-called green valley (i.e. the intermediate area between the regions including most red and elliptical galaxies and most blue and spiral galaxies, known as red sequence and blue cloud, respectively) in the colour-magnitude diagram. The occurrence of quenched disks and red spiral galaxies in the inner regions of galaxy clusters suggests that morphological transformations are delayed with respect to the decline in star formation. In their extensive analysis of SDSS and GALEX low-redshift data \citet{schawinski14} find that the quenching of star formation in late type galaxies is a gradual process which takes place on timescales $>$1 Gyr. In addition the authors argue that the quenching process does not necessarily involve or result in a morphological transformation and hence generates red disk galaxies.\\

In this paper we present a spectroscopical and, for the first time, morphological analysis of the galaxies in the merging cluster Abell 2384. We exploit new 2dF spectral data complemented with already existing EFOSC2 spectra and ESO Wide Field Imager (WFI) R-band and B-band imaging.\\
Abell 2384 is classified as an Abell cluster of richness 1 and BM-type II-III at z$\simeq$0.094. X-ray studies (EINSTEIN, \citealt{ulmer82}, ROSAT, \citealt{henriksen96, degrandi99}, Chandra, \citealt{markevitch02}) and Palomar Schmidt 1.5m \citep{oegerle87} observations established a bimodal ICM and galaxy distribution. \citet{cypriano04} performed a weak-lensing analysis on the main cluster of the Abell 2384 system. Analyses based on XMM-Newton and ESO Wide Field Imager observations confirmed bimodality and support a scenario in which Abell 2384 represents a post merger between a main cluster in the north and a less massive subcluster in the south (\citealt{maurogordato11}).\\
   
Throughout this paper we assume $H_{0}=70$ km/s/Mpc, $\Omega_{m}=0.3$, $\Omega_{\Lambda}=0.7$. At the systemic cluster redshift of 0.094, 1 arcmin corresponds to $\sim$105 kpc in this cosmology. 

\section{The data}
Using the Two Degree Field (2dF) system on the AAT \citep{lewis02} multifibre spectroscopical observations were carried out in the central 2x2 deg$^{2}$ of the galaxy cluster Abell 2384 in June 2005. We used the 400 fibre positioning system and the 2dF double spectrograph to carry out two sets of observations both centred at $\alpha$=$21^{h}52^{m}21.96^{s}$, $\delta$=$-19^{\circ}32^{m}48.65^{s}$. Fibre allocation was performed using the 2dF \texttt{Configure} programme. Considering a set of instrumental limitations such as minimum distance between allocated fibres this software optimises fibre placement based on user-defined weights associated to each target present in the input catalogue. The required input file was created on the basis of apparent magnitudes in our ESO WFI deep imaging data \citep{maurogordato11}, completed by the SuperCOSMOS catalogue \citep{hambly01} for the regions not covered by WFI observations. \\ 
We used the same 2dF grating (300B) in the two available spectrographs of 2dF and we adopted a central wavelength of 5806 $\mathring{A}$. Our observations thus gave 400 spectra per observing run and covered the approximate wavelength range 3800-8200 $\mathring{A}$. At A2384 redshift (z$\simeq$0.094) this wavelength interval includes all spectral features from [OII] 3727 $\mathring{A}$ to H$\alpha$ 6563 $\mathring{A}$. Two runs of observations were carried out in June 2005 with a total exposure time of 3600 seconds per run, divided in two exposures of 1800 seconds in order to eliminate cosmic rays. The data were reduced using the 2dF data reduction pipeline software \texttt{2dfdr}.\\
For details on our WFI deep imaging observations and on our complementary EFOSC2 spectral data see \citet{maurogordato11}. 

\subsection{Redshift determination}
Redshift determination for our new spectral data was carried out using the automatic redshift codes for 2dF and AAOmega spectra \texttt{runz} \citep{colless01} and \texttt{autoz} \citep{baldry14}. Errors were estimated comparing the redshifts of 177 objects that had been observed twice \citep{milvang-jensen08} in the course of our 2dF spectroscopy. We found the best results of both codes to agree well within the typical redshift error of $\delta_{z}$=0.0002 (estimated via biweight statistics, see \citealt{beers90} for details). For all of our further analyses we used the redshift results associated with the highest level of confidence in the output of the respective code. All of these redshifts also underwent a visual check.\\
Among the total number of 672 non-blank sky spectra 177 objects were observed twice. The remaining set of 495 single spectra contains 130 objects which we identify as stars. Within the 365 galaxy spectra we detect 22 spectra to be corrupted and hence not usable for further analysis. Using redshift determination quality as a final separation criterion we find 38 of the 343 redshifts of non-corrupted galaxy spectra to be unreliable, i.e. being associated with a confidence level of less than 80$\%$. The confidence level is defined via the ratio of the first and the subsequent peaks in the cross-correlation function implemented in the redshift algorithms (see \citealt{colless01} and  \citealt{baldry14} for details). Our reliable 2dF redshift sample finally consists of 305 objects of which 229 reach a confidence level of greater than 90$\%$ and 76 redshifts lie between 80$\%$ and 90$\%$. All of these 305 galaxies passed the visual redshift check.\\
We complement our redshift sample with 68 objects from the catalogue of \citet{maurogordato11} (EFOSC2 spectra) with z$>$0.005. Five of these galaxies are also part of our 2dF redshift sample and we find that the measured redshifts are in compliance within the errors. For these five cases we keep the 2dF redshift values because of the wider spectral range compared to EFOSC2 and the therefore more robust redshift estimates. We hence have at hand a sample of 368 reliable redshifts in the two-degree-field around Abell 2384.  

    \begin{figure}
   \centering
   \includegraphics[angle=270,width=\columnwidth]{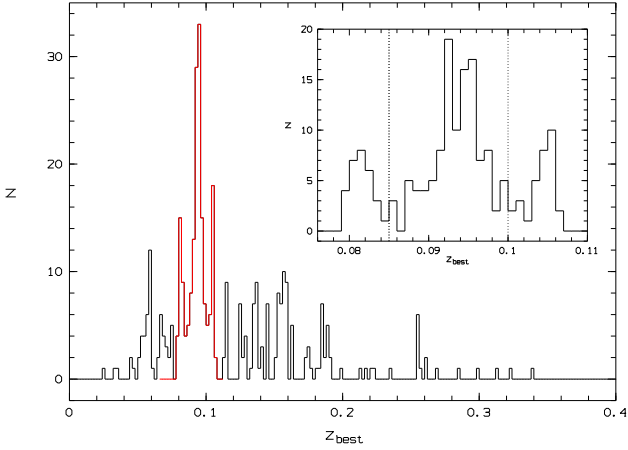}
      \caption{Redshift histogram of the 368 galaxies with reliable redshifts in the 2x2 square degrees field centred on A2384. The binsize is $\Delta$z=0.002. \textit{Inlay:} zoom on the red coloured part of the total distribution (0.0796$\leq$z$\leq$0.1065, $\Delta$z=0.001). The dashed lines indicate the redshift limits for cluster members found with EMMIX (see text for details).}
         \label{fig:hist1_red_inlay}
   \end{figure}     

\section{Redshift distribution and cluster membership}
\begin{figure}
   \centering
   \includegraphics[angle=0,width=\columnwidth]{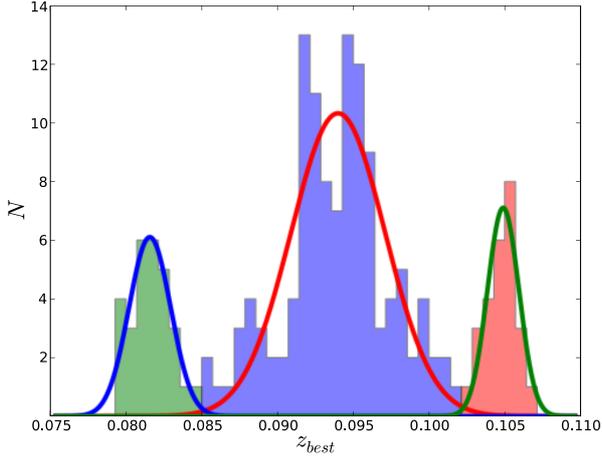}
      \caption{Redshift histogram of 173 galaxies in the range 0.0796$\leq$z$\leq$0.1065 with a binsize of $\Delta$z=0.0007. The different colours correspond to the partitions found by EMMIX (partition 1, 2 and 3 from left to right). The solid lines indicate the Gaussian fits to the respective partition. Note that for the sake of visibility the colours have been permuted between the histograms and the fits.}
         \label{fig:173redshifts}
   \end{figure}

\begin{figure}
   \centering
   \includegraphics[angle=0,width=\columnwidth]{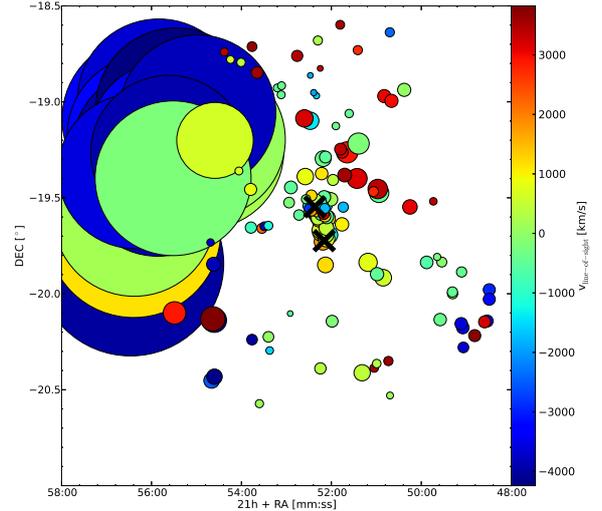}
      \caption{DS-test results for the clustersample of 173 galaxies (after application of $ \pm $5000 km/s line-of-sight velocity cuts). Their spatial locations are indicated by circles with radii proportional to $e^{\delta}$ where $\delta$ is the DS-test measure of the local deviation from the global velocity dispersion and mean recessional velocity, i.e. larger symbols correspond to a higher significance of substructure. The colours indicate the rest-frame velocity relative to the systemic cluster line-of-sight velocity in km/s. The inner cluster regions are identifiable as the overdense central area. The positions of both brightest cluster galaxies (BCG1 and BCG2) are shown by black crosses. The 3D-substructure detected in the eastern region of the cluster consists entirely of galaxies assigned to partition 1 in Fig. \ref{fig:173redshifts}}
         \label{fig:DS173_scaled}
   \end{figure}

In Fig. \ref{fig:hist1_red_inlay} we show the redshift distribution of all 368 galaxies with reliable redshift in the 2x2 square degrees field centred on A2384. There is a clear concentration of galaxies between z$ \simeq $0.08 and z$ \simeq $0.11. The distribution reaches its maximum at a redshift of z$ \simeq $0.0942. These findings are in agreement with earlier results (e.g. \citealt{markevitch02, cavagnolo09}). Aiming at a definition for cluster membership we first narrow down our sample to galaxies within $ \pm $5000 km/s around the line-of-sight velocity corresponding to z=0.0942. This common redshift filter already excludes most background and foreground field galaxies \citep{boesch13}. We are left with 173 galaxies in the redshift range 0.0796$\leq$z$\leq$0.1065 (red coloured bins and inlay in Fig. \ref{fig:hist1_red_inlay}). Next we run the EMMIX software \citep{mclachlan99} on the redshift distribution of this subsample. This well-established realisation of an expectation-maximisation algorithm fits the redshift distribution with mixtures of one to 10 Gaussian distributions, not requiring any initial guess for data subclustering. The EMMIX output provides three criteria for the number of sub-components (also referred to as partitions) found in the data. These are the Bayesian information criterion (BIC), the Akaike information criterion (AIC) and the approximate weight of evidence (AWE). The most probable number of partitions is the one for which the respective criterion value is minimal. For galaxy clusters the best results are obtained using the BIC (Rostagni 2013, priv. comm.). As a test we applied EMMIX to our 2dF-data on Abell 3921 for which subclustering in redshift space could already be investigated using other techniques (see \citealt{pranger13} for details). The test outcome is in compliance with the validity of the BIC. In addition to the computation of the different partitions and the associated criteria values, EMMIX performs a bootstrap analysis which sequentially compares the different numbers of partitions (1 vs. 2, 2 vs. 3 etc.) and returns a P-value. The P-value is a measure of significance and mathematically represents the probability for the found data configuration to appear by chance (under the given null hypothesis). This means that a small P-value suggests the rejection of the null hypothesis. The smaller the P-value, the more significant the result (i.e. the deviation from the null hypothesis).\\
According to the BIC (and AWE) the best fit to the A2384 data is a superposition of three Gaussian distributions shown in Fig. \ref{fig:173redshifts}. Moreover, the bootstrap analysis returns the smallest significance value (P$<$3$\cdot10^{-3}$) for the case of three partitions. Properties calculated from the underlying data (using biweight statistics for $S_{BI}$, i.e. velocity dispersion) are listed in Table \ref{table:tab1}.\\ 
Motivated by the striking trimodality in the redshift distribution and its confirmation by EMMIX we apply the Dressler-Shectman (DS-)test \citep{dressler88} to the subsample of 173 galaxies. An illustration of this test's outcome is given in Fig. \ref{fig:DS173_scaled}. It reveals some 3D-substructure in the eastern part of the field-of-view with a significance value P=0 for $10^{6}$ Monte-Carlo iterations. This means that none of the $10^{6}$ random re-configurations of the galaxies on the sky plane results in a subclustering indicator $\Sigma\delta$ (see caption of Fig. \ref{fig:DS173_scaled}) greater than the one calculated from the original galaxy positions. The DS-test highlights a group of 21 galaxies which separates into six objects with redshifts roughly around the systemic cluster redshift (0.0931$\leq$z$\leq$0.0976) and 15 objects with lower redshifts in the narrower range 0.0796$\leq$z$\leq$0.0830. The latter constitute more than 50$\%$ of the galaxies in partition 1 of the EMMIX output (see Fig. \ref{fig:173redshifts}) and show a median redshift of $<$z$>$=0.0810 and a velocity dispersion of $351^{+53}_{-135}$ km/s. We interpret these 15 galaxies as (part of) a massive group falling onto the cluster from the east (projected onto the plane of the sky) with a high peculiar line-of-sight velocity towards the observer.\\
Apart from this eastern group the DS-test does not reveal any further substructure in the subsample of 173 galaxies. In particular it does not detect a high-redshift counterpart to the low-redshift eastern group that could be associated with partition 3 of the EMMIX results. However, the high redshift objects appear concentrated in the north-west of the field-of-view rather than randomly distributed (see Fig. \ref{fig:DS173_scaled}).\\
Based on the EMMIX and DS-test output we define the sample of 118 galaxies assigned to partition 2 as our preliminary cluster sample. To define cluster membership more precisely we apply the standard iterative 3$\sigma$ clipping \citep{yahil77} and a modified 3$\sigma$ clipping involving the biweight estimators for location and scale \citep{beers90} to the preliminary sample. Neither version of the 3$\sigma$ clipping leads to sample reduction. Hence we adopt the preliminary cluster sample as it stands, i.e. 118 galaxies constituting partition 2 of the EMMIX output. We will refer to it as the total cluster sample (TCS) in the following.\\
       
\begin{table}    
\centering                          
\begin{tabular}{c c c c c}        
\hline\hline                 
partition & $N_{gal}$ & $<$z$>$ & $S_{BI}$ & Allocation rate \\&&&[km/s]\\    
\hline            
\noalign{\smallskip}   
   1 & $29$ & $0.0815$ & $387^{+45}_{-90}$ & $0.994$\\      
\noalign{\smallskip}  
   2 & $118$ & $0.0941$ & $1051^{+96}_{-132}$ & $0.982$\\
\noalign{\smallskip}    
   3 & $26$ & $0.1049$ & $301^{+38}_{-72}$ & $0.998$\\
\noalign{\smallskip}     
\hline                                   
\end{tabular}
\caption{Number of galaxies per partition ($N_{gal}$), median redshift ($<$z$>$), velocity dispersion ($S_{BI}$) and allocation rate for the three partitions found by EMMIX. The allocation rate is a relative estimate for the reliability of the assignment of galaxies to their host partition.
 }             
\label{table:tab1} 
\end{table}       
   
\subsection{Cluster substructure}
\label{susec:clusub}
\begin{figure}
   \centering
   \includegraphics[angle=0,width=\columnwidth]{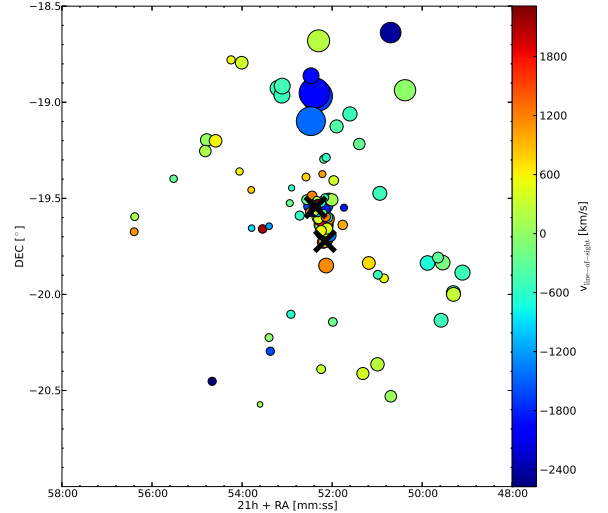}
      \caption{Same as Fig. \ref{fig:DS173_scaled} but for the total cluster sample (TCS). The brightest cluster galaxies are marked by black crosses. The DS-test detects tentative substructure in the northern and western cluster outskirts.}
         \label{fig:DS_1}
   \end{figure}

\begin{figure}
   \centering
   \includegraphics[angle=0,width=\columnwidth]{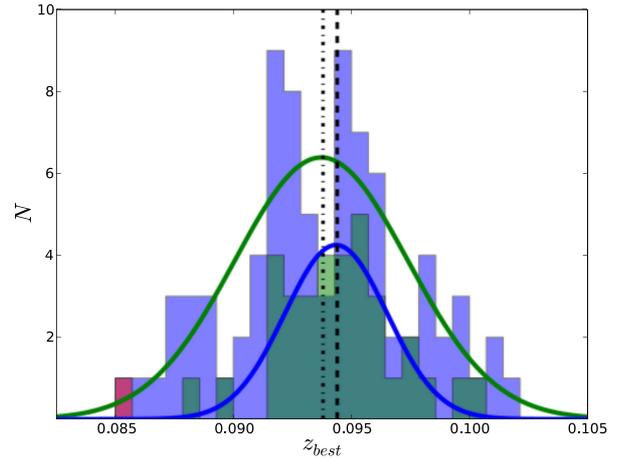}
      \caption{Redshift histogram of the northern cluster sample (NCS, 83 galaxies) in blue and of the southern cluster sample (SCS, 34 galaxies) in green (binsize $\Delta$z=0.0007). The object which gets removed from the SCS after the 3$\sigma$ clippings is shown in red. The solid lines indicate Gaussian fits to the respective cluster subsample. Note that for the sake of clarity the colours have been interchanged between the histograms and the fits. The dash-dotted and dashed line show the median redshift of the NCS and the SCS, respectively.}
         \label{fig:n83s35_red}
   \end{figure}

\begin{figure}
   \centering
   \includegraphics[angle=0,width=\columnwidth]{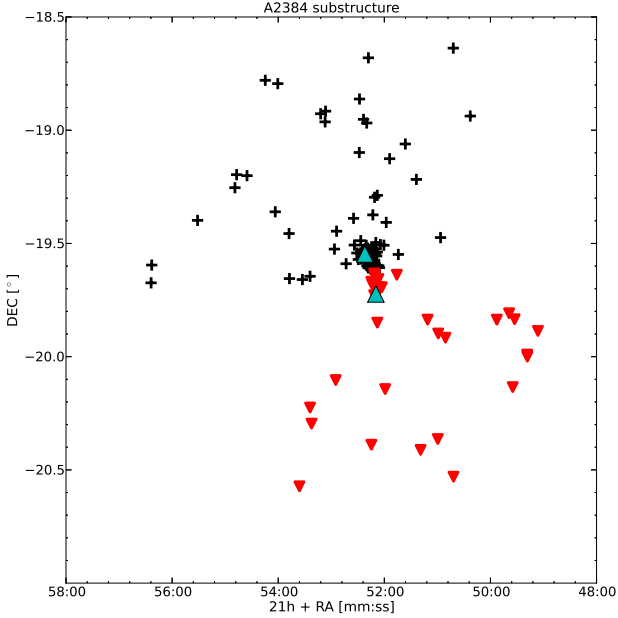}
      \caption{Substructure of A2384. Black crosses represent galaxies in the northern subluster, red triangles indicate objects belonging to the southern subcluster. BCG1 and BCG2 are indicated by cyan triangles.}
         \label{fig:N_S}
   \end{figure}

Previous investigations on the central $\sim$3x3 Mpc$^{2}$ have shown that A2384 represents a merging cluster system \citep{ulmer82, west95, henriksen96, maurogordato11} with a more massive main cluster in the north and a smaller subcluster in the south (mass ratio $\sim$2.3:1). Both clusters are associated with giant elliptical galaxies (BCG1, BCG2) the latter of which resides at the location of the corresponding peak in X-ray emission. BCG1 is slightly ($\sim$15 arcseconds, corresponds to $\sim$25 kpc at the cluster redshift) offset from the northern X-ray maximum \citep{maurogordato11}. To further investigate the substructure of A2384 on the basis of our combined 2dF and EFOSC2 data we apply the DS-test also to the TCS. The resulting significance value P$<$0.05 for $10^{6}$ Monte-Carlo iterations is clearly below the generally adopted threshold of P=0.1 and hence indicates 3D-substructure within the TCS. The graphical DS-test output presented in Fig. \ref{fig:DS_1} shows some substructure residing in the northern and western cluster outskirts. In particular the DS-test does not allow to separate the TCS into a northern and a southern subcluster. This is mainly due to the similar redshifts of the subclusters. Having no other separation criterion at hand we use the projected distance of each galaxy to BCG1 and BCG2, respectively, to split the TCS into a northern cluster sample (NCS, 83 galaxies) and a southern cluster sample (SCS, 35 galaxies initially). After reapplication of the 3$\sigma$ clippings to the cluster subsamples the galaxy number in the SCS is reduced by one object whereas the NCS is not affected. The redshift and spatial distributions of both cluster subsamples are depicted in Figs. \ref{fig:n83s35_red} and \ref{fig:N_S}, respectively.\\
Follow-up EMMIX runs on both cluster subsamples indicate unimodality (in spite of the two-peaked redshift distribution of the NCS), follow-up DS-tests show significance values of P$\sim$0.10 for both the NCS and the SCS, using $10^{6}$ Monte-Carlo iterations. This implies that the tentative 3D-substructures detected in the outskirts of the TCS become insignificant when separately analysing the cluster subsamples.\\
In Table \ref{table:tab2} we list the median redshift and the biweight estimators for scale ($S_{BI}$, corresponding to velocity dispersion) as well as estimates for $r_{200}$ and total virial mass calculated on the assumption of isothermal spheres (\citealt{carlberg97}) and on the basis of 3D harmonic radius ($r_{h}$, \citealt{small98}) and velocity dispersion. The mass and radii estimators are given by:

\begin{equation}
      M = \frac{3\pi S_{BI}^{2}r_{h}}{G},
      \label{eq:m}
\end{equation} 

\begin{equation}
      r_{200} = \frac{\sqrt{3}S_{BI}}{10H_{0}\sqrt{\Omega_{m}(1+z)^{3}+\Omega_{\Lambda}}}
      \label{eq:r200}
\end{equation}

and

\begin{equation}
      r_{h} = \frac{2}{N(N-1)D\left( \sum_{i}\sum_{j<i}\frac{1}{\Theta_{ij}}\right) ^{-1}},
      \label{eq:rh}
\end{equation}

respectively, where $G$ is the gravitational constant, $N$ is the number of galaxies, $D$ is the radial distance to the cluster and $\Theta$ is the angular distance between galaxies $i$ and $j$.\\
Our biweight estimates for velocity dispersion of the TCS and the NCS are in excellent agreement with \citet{maurogordato11}. Our estimate for the velocity dispersion of the SCS is, however, $\sim$160 km/s lower than their result. \citet{maurogordato11} find that both the northern and southern subcluster are above the galaxy velocity dispersion vs. X-ray ICM temperature ($\sigma - T_{x}$) relation (see \citealt{wu98}) which indicates that they are dynamically perturbed and non-virialized. For the NCS this is also the case with our data. It is not the case for the SCS which, thanks to the better coverage, is in perfect agreement with the $\sigma - T_{x}$ relation.\\
However, considering the deviation of the NCS from the $\sigma - T_{x}$ relation and the dynamically active state of A2384 in general we conclude that our results for velocity dispersion and accordingly also for $r_{200}$ and $M$ are overestimated for the TCS and the NCS. The mean harmonic radius for the SCS is larger than for the NCS and similar to the TCS. Compared to our overestimated values for $r_{200}$ we expect the mean harmonic radius to be significantly smaller which is true for the TCS and the NCS but not for the SCS. This is due to the estimation technique and the still relatively small sample size of the SCS. As a consequence also the mass of the southern subcluster is probably overestimated. Since we have to assume that all three sample masses are overestimated (due to overestimated velocity dispersion or overestimated mean harmonic radius, respectively) we regard our result of $2.34^{+0.08}_{-0.07} \cdot 10^{15} M_{\odot}$ for total cluster mass as an upper limit. This value is larger by a factor of 1.17 than the upper limit to total cluster mass found by \citet{maurogordato11}, however, it is still in agreement with the order of magnitude of their mass estimate. Given that velocity dispersion enters squared in the virial mass term whereas the mean harmonic radius enters only linearly (see \citealt{carlberg97}) we interpret the subcluster mass ratio of 1.6:1 following from our mass estimates as an upper limit, too. Consequently also the mismatch between the mass estimate for the TCS and the sum of both subcluster masses would tend to increase for corrected masses. We cannot overcome these limitations since we cannot separate the northern and southern subcluster in 3-D physical space.

\begin{table*}    
\centering                          
\begin{tabular}{c c c c c c c}       
\hline\hline                 
Subsample & $N_{gal}$ & $<$z$>$ & $S_{BI}$ & $r_{200}$ & $r_{h}$ & $M$ \\&&&[km/s]&[kpc]&[kpc]&[$10^{15} M_{\odot}$]\\    
\hline            
\noalign{\smallskip}   
TCS & $118$ & $0.0939$ & $1051^{+96}_{-132}$ & $2488^{+213}_{-301}$ & $1522^{+49}_{-48}$ & $2.34^{+0.08}_{-0.07}$\\      
\noalign{\smallskip}  
NCS & $83$ & $0.0937$ & $1134^{+99}_{-163}$ & $2684^{+234}_{-386}$ & $1026^{+42}_{-40}$ & $1.84^{+0.08}_{-0.07}$\\
\noalign{\smallskip}    
SCS & $34$ & $0.0944$ & $743^{+119}_{-211}$ & $1761^{+282}_{-499}$ & $1516^{+208}_{-169}$ & $1.17^{+0.16}_{-0.13}$\\
\noalign{\smallskip}    
\hline                                   
\end{tabular}
\caption{Number of member galaxies, median redshift, velocity dispersion ($S_{BI}$), $r_{200}$, $r_{h}$ and $M$ for the total cluster sample and both subsamples.
 }             
\label{table:tab2} 
\end{table*}
           
\subsection{Field sample definition}
Since our WFI imaging on A2384 only covers the inner 30x30 arcminutes we have photometric and morphological information on only 24 field galaxies, defined as no members of the TCS. This small field sample strongly differs from the cluster sample in its redshift distribution. If, to allow a fair comparison, we apply upper and lower redshift limits to the 24 galaxies such that their mean and median redshift becomes similar to the TCS we end up with 14 galaxies in the remaining field sample. We consider this too few for robust statistical analysis. We hence make use of the field sample defined in our analysis of the galaxy population of Abell 3921 \citep{pranger13}. This choice is justified by the fact that Abell 3921 is located at a similar redshift (z$\simeq$0.093) as Abell 2384 (z$\simeq$0.094). In addition, our optical images of A3921 and A2384 have been taken with the same telescope, instrument (ESO WFI) and filter and under comparable seeing conditions.\\
The original A3921 field sample consists of 83 galaxies. To adjust it to the A2384 redshift distribution we remove two low-redshift objects and thus end up with 81 galaxies in the A2384 field sample. This basic field sample will be used whenever we present comparisons between the field and either the TCS, NCS or SCS, respectively. For morphological analyses the A3921 field sample had to be reduced to 74 galaxies due to image defects. We also adopt this basic morphological field sample and find that it does not require any alteration in order to optimally match our corresponding morphological cluster sample in redshift- and magnitude distribution. 

\section{Spectroscopical analysis}

\subsection{[OII] and H$\alpha$ equivalent width measurement}
\label{susec:minew}
We determined emission line equivalent widths (EWs) from our 2dF and EFOSC2 spectra in the same way as in \citet{pranger13}. For 23 objects with low S/N ratios, the [OII] line could not be fitted with a doublet but only with a single line profile. We corrected the [OII] EWs of these galaxies with a factor of 1.09 determined from 25 objects with the highest S/N, for which we compared robust doublet fits to single fits.\\
We also estimated the EW upper limits of [OII] and H$\alpha$ analogously to our spectral analysis of galaxies in Abell 3921 using mock spectra (see \citealt{pranger13} for details). We found a minimum EW of 5.0 $\mathring{A}$ for [OII] and 1.9 $\mathring{A}$ for H$\alpha$. Note that due to the narrower spectral range we could not measure H$\alpha$ EWs for the EFOSC2 spectra in our sample.       

          \begin{figure*}
   \centering
   \includegraphics[angle=0,width=\textwidth]{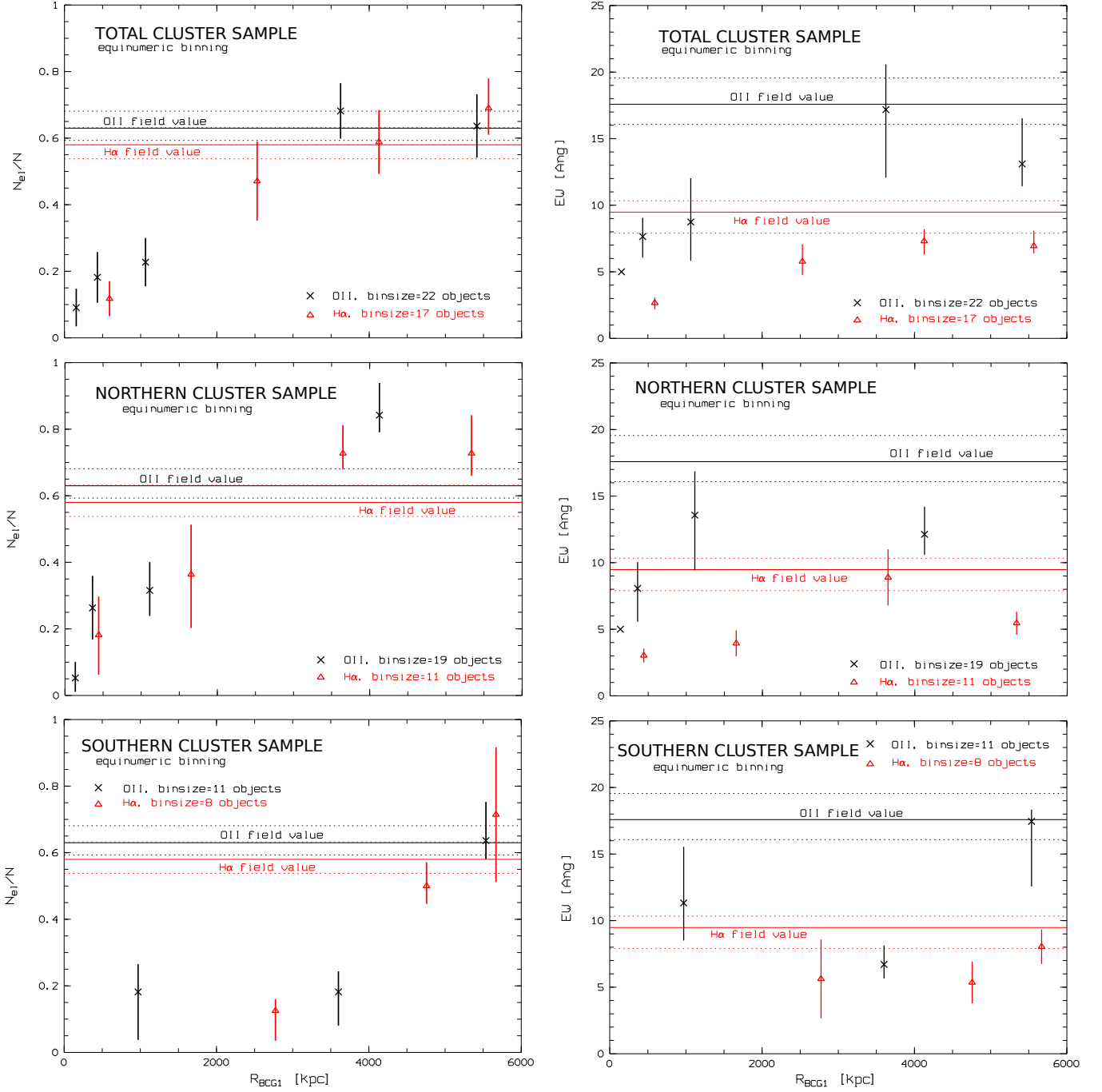}
      \caption{\textit{Left column:} Fraction of galaxies showing [OII] or H$\alpha$ emission lines, respectively, vs. distance from the cluster centre. \textit{Right column:} Mean equivalent width of [OII] and H$\alpha$ emission lines for all objects (objects without detectable lines get assigned a lower threshold
value) vs. distance from the cluster centre. \textit{Top to bottom:} Total cluster sample (TCS), northern cluster sample (NCS) and southern cluster sample (SCS). Note that for the southern cluster sample the increasing trend only sets in at clustercentric distances $>$3 Mpc and does not reach the field level before $R_{BCG1}\simeq$6 Mpc.}
         \label{fig:spec}
   \end{figure*}

\subsection{Equivalent widths and fractions of EL galaxies}
\label{susec:radial}
Fig. \ref{fig:spec} shows equivalent widths of [OII] and H$\alpha$ lines as well as the fraction of EL-galaxies (i.e. galaxies definitively showing [OII], H$\alpha$ or both emission lines) as functions of clustercentric distance (the cluster centre taken to be coincident with BCG1), for the TCS, NCS and SCS, respectively. For the right hand side panels (i.e. equivalent width plots), we assigned an equivalent width of 5.0 $\mathring{A}$ to objects without a detectable [OII] doublet, and an EW of 1.9 $\mathring{A}$ to the objects without a detectable H$\alpha$ line (see Sec. \ref{susec:minew}).\\   
All errors given in Fig. \ref{fig:spec} are generated via bootstrapping. The clustercentric distances ($R_{BCG1}$) are median values for each bin. To rule out potential binning biases we analysed different binning methods (equidistant and equinumeric binning) and a wide range of binsizes. The plots shown here reflect trends that are robust against binning changes. As in \citet{pranger13} we use an equinumeric binning in these plots, i.e. whenever we show a given quantity as a function of radius, we keep the number of galaxies per bin fixed throughout the plot, except for the last data point which contains the respective remainder according to the total number of objects within the sample under investigation. Horizontal lines indicate the respective field values.\\
Note that due to technical problems object-spectra correlation was not possible for eight galaxies in the EFOSC2 dataset which leaves us with a spectroscopical sample of 110 galaxies. Further we point out again that due to the narrow spectral range we do not have H$\alpha$ measurements for the EFOSC2 spectra. Thus the sample size decreases to 67 in the case of H$\alpha$ analyses. In effect, in all panels of Fig. \ref{fig:spec} the first data point for H$\alpha$ is shifted outwards with respect to its [OII] counterpart.\\ 
It is noticeable that for the TCS and the NCS both investigated quantities show an increase towards larger clustercentric radii. These trends become stagnant when the field levels are reached (at clustercentric distances $>$3.5 Mpc). This result is in compliance with previous studies on the dependence of galaxy properties on environment (e.g. \citealt{verdugo08, koyama13}). However, this is only partly the case for the SCS. The first data point appears $\sim$600 kpc (i.e. half of the projected distance between BCG1 and BCG2) further outwards in the SCS plots. This is due to the distance measurement which was taken with respect to BCG1. This is motivated by the greater mass of the northern subcluster and the resulting expectation that environmental effects on the galaxy properties could be best traced as functions of distance to the gravitational cluster centre. Moreover, the increasing trend only sets in at clustercentric distances $>$3 Mpc and does not reach the field level before $R_{BCG1}\simeq$6 Mpc.\\         
  
\section{Morphological analysis}
\label{sec:morph} 
\begin{figure}
   \centering
   \includegraphics[angle=90,width=\columnwidth]{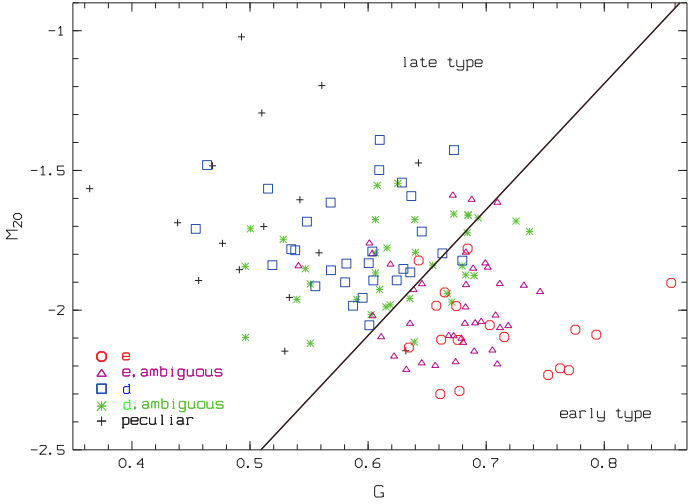}
      \caption{Overall morphological sample (136 objects) in Gini-M$_{20}$ space, morphologically subdivided into ellipticals (red circles and purple triangles), disks (blue squares and green stars) and peculiars (black crosses). The black line represents the cut applied to the unambiguously classified galaxies (blue squares and red circles) finally dividing the whole sample into late type galaxies (upper left corner) and early type galaxies (lower right corner). Objects classified as peculiar are included in the respective subsample. Note that we use this quantitative criterion throughout our morphological analysis.}
         \label{fig:GiniM20}
   \end{figure}

\begin{figure*}
   \centering
   \includegraphics[angle=0,width=\textwidth]{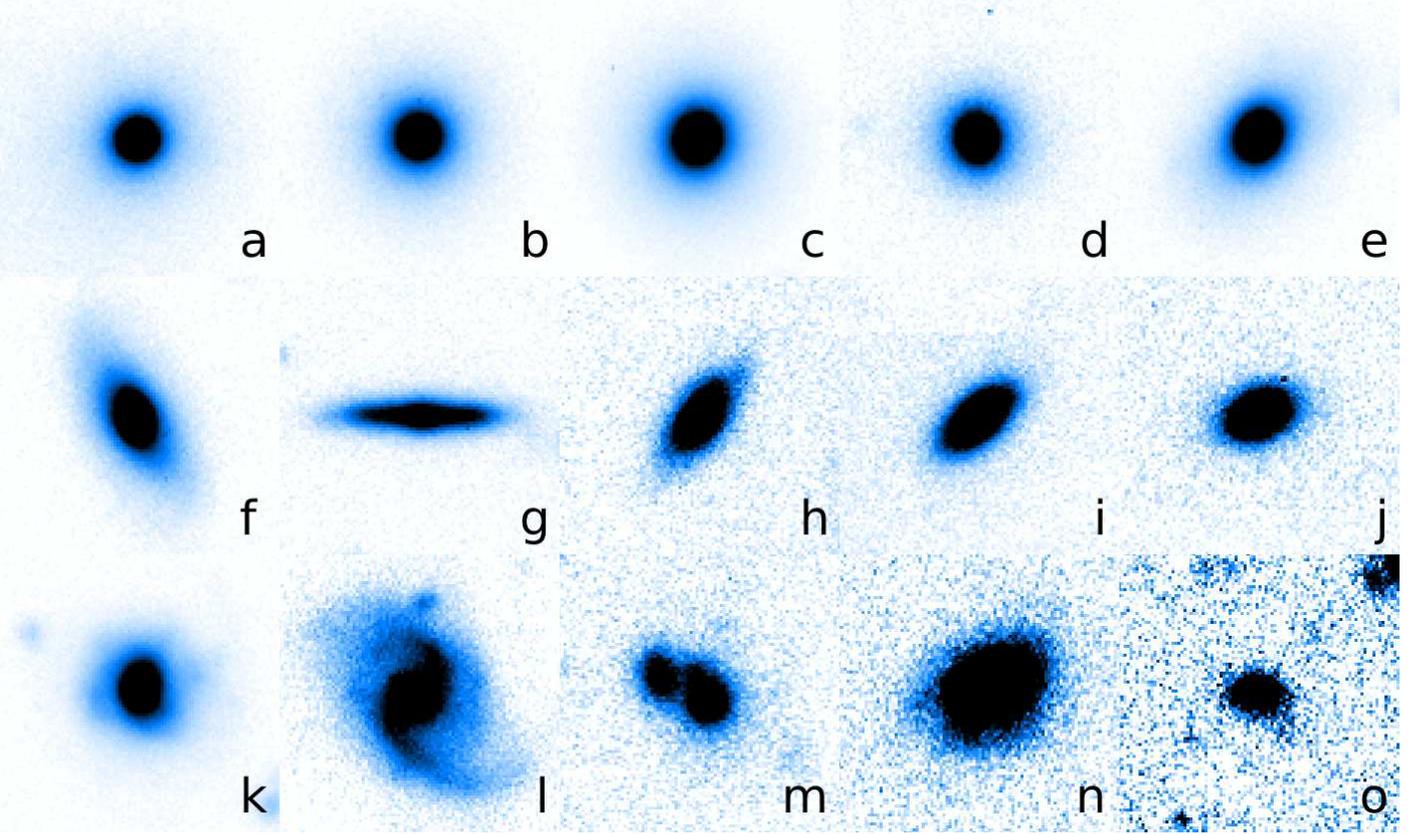}
      \caption{WFI R-band images (22x22 arcseconds each) of typical early type cluster galaxies (top row, images a-e) and typical late type cluster galaxies (middle row, images f-j) according to Gini/M$_{20}$ classification. The bottom row (images k-o) shows the peculiar cluster sample. We use logarithmic contrast cuts, normalised to the mean intensity of each image.}
         \label{fig:10_2b}
   \end{figure*}

\begin{figure}
   \centering
   \includegraphics[angle=0,width=\columnwidth]{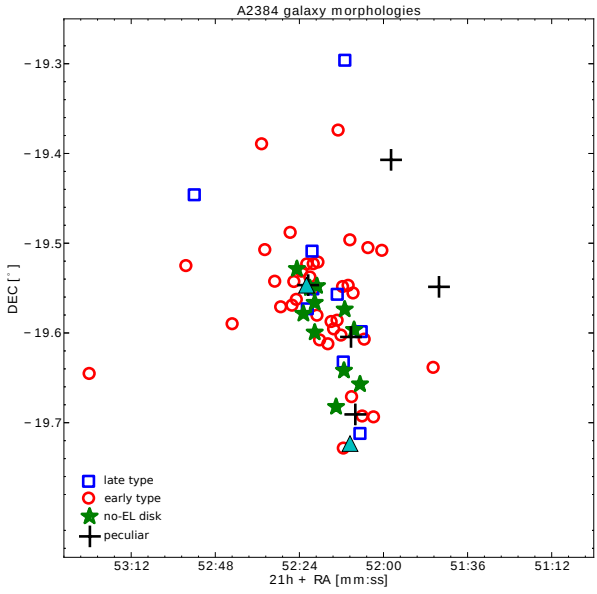}
      \caption{Spatial distribution of the MCS. Early type objects are shown as red circles, late type galaxies are represented by blue squares. No-EL disks are shown separately as green stars. Black crosses indicate morphologically distorted (i.e. peculiar) galaxies. BCG1 and BCG2 are represented by cyan triangles.}
         \label{fig:morphs}
   \end{figure}

\begin{figure}
   \centering
   \includegraphics[angle=0,width=\columnwidth]{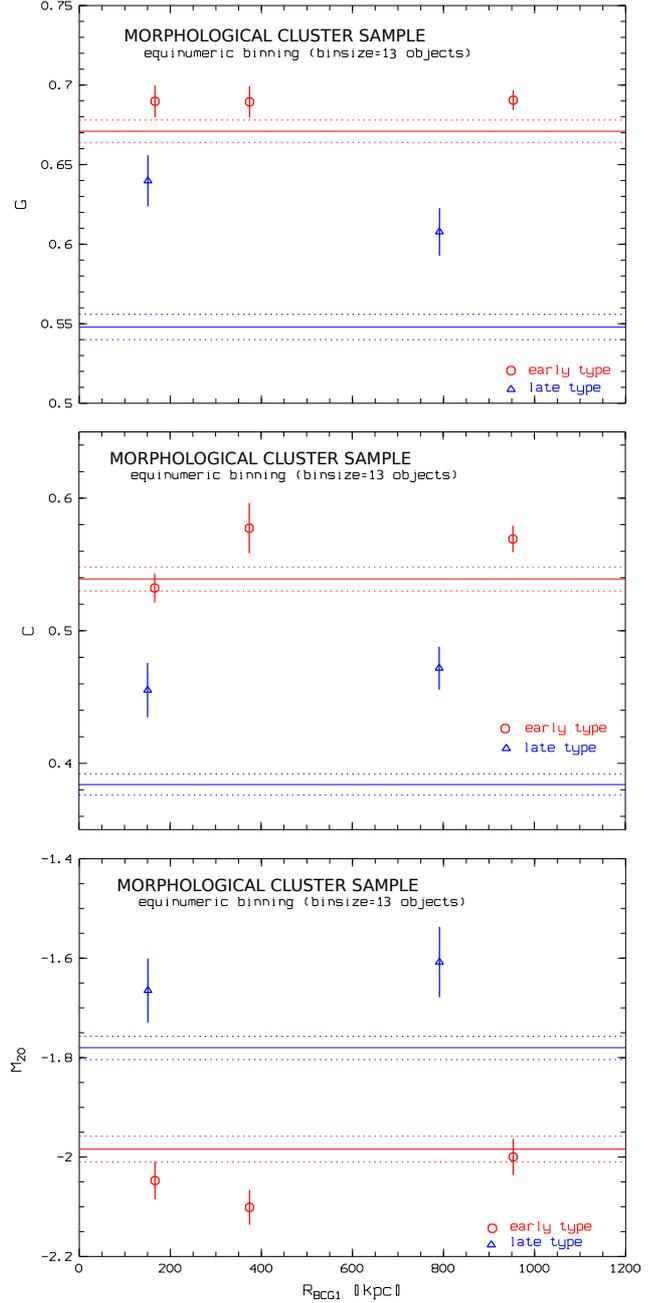}
      \caption{\textit{Top to bottom:} Gini coefficient (G), concentration index (C) and second order moment of the brightest 20$\%$ of the galaxy (M$_{20}$) as a function of clustercentric distance for early type galaxies (red circles) and late type galaxies (blue triangles). Data points represent median values, errors are estimated via bootstrapping.}
         \label{fig:morph_r}
   \end{figure}

For our morphological analysis we use the WFI imaging data of the inner 30x30 arcminutes of A2384 by \citet{maurogordato11}. We have photometric data for 62 cluster member galaxies. In the following, we refer to this sample as our morphological cluster sample (MCS). As in \citet{pranger13} we calculate Gini coefficient (G), concentration index (C) and M$_{20}$ index for each of these galaxies. We also visually classify each galaxy as spiral (21), elliptical (36) or peculiar, i.e. showing morphological distortions (5). Combining the MCS with our morphological field sample (74 galaxies) we end up with a combined morphological sample of 136 objects.\\ 
Using only cases of \textit{unambiguous visual classifications} (34$\%$ of our sample) we define a dividing line in Gini-M20 space such that the separation of late type (left-hand side) and early type (right-hand side) galaxies is optimised. This line is then held fixed and used for the classification of the \textit{whole} sample. We will use this \textit{purely quantitative} morphological classification in the following analysis. The peculiar galaxies show a wide spread in Gini-M$_{20}$ space (see Fig. \ref{fig:GiniM20}) which makes it impossible to define a quantitative Gini-M$_{20}$ criterion for peculiarity. The reason for this is the limited spatial resolution of our ground-based imaging. Thus, in contrast to the distinction into late type and early type galaxies, we keep a visual classification for the peculiars \textit{only}. Fig. \ref{fig:GiniM20} shows the combined morphological sample (cluster and field) in Gini-M$_{20}$ space. Fig. \ref{fig:10_2b} shows WFI R-band images of typical early type, late type and peculiar cluster members. In Fig. \ref{fig:morphs} the positions and morphological types of all 62 members of the morphological cluster sample are depicted. It clearly shows the north-south elongated structure of the cluster core regions. Although the majority of galaxies in these regions are early type objects, it is noticeable that they also host a relatively large number of late type galaxies. In particular we find 11 no-EL disks and three peculiar late type galaxies (see Sec. \ref{susec:nels}) residing in the densest regions around the BGCs. Note that four out of five peculiar cluster galaxies are late type galaxies according to the separation in Gini/M$_{20}$ space. Fig. \ref{fig:morph_r} illustrates morphological quantities as functions of clustercentric distance for the MCS (i.e. cluster member galaxies within the inner 30x30 arcminutes, centred on BCG1) split up into early type and late type galaxies.\\
    
\subsection{No-EL disks and peculiar galaxies}
\label{susec:nels}
\begin{figure}
   \centering
   \includegraphics[angle=270,width=\columnwidth]{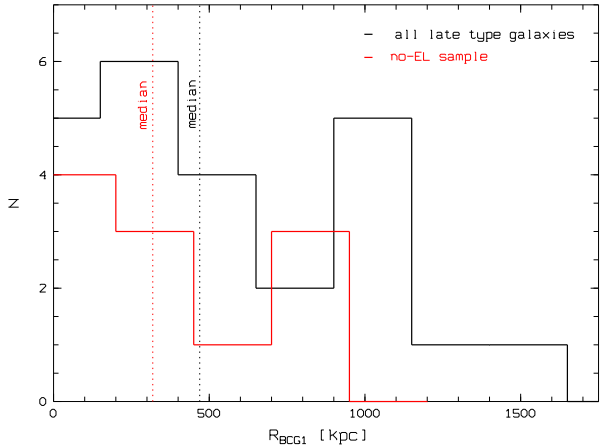}
      \caption{Number of morphologically classified late type galaxies as a function of clustercentric distance. The no-EL disk sample is found to be on average closer to the cluster centre than the sample containing all late type galaxies.}
         \label{fig:nelshist}
   \end{figure}

\begin{figure}
   \centering
   \includegraphics[angle=0,width=\columnwidth]{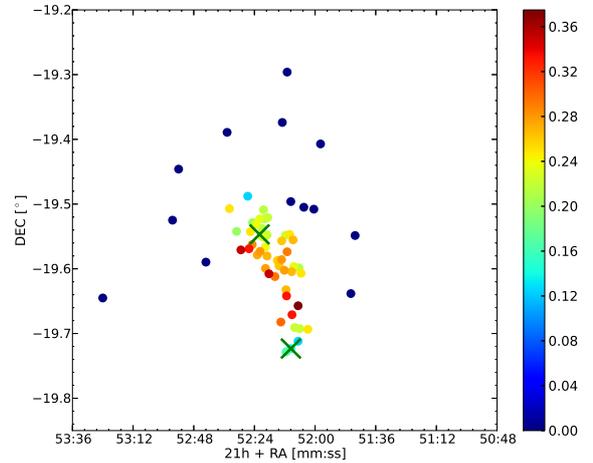}
      \caption{Local fractions of no-EL disk galaxies (with respect to the MCS). BCG1 and BCG2 are represented by green crosses. 
      }
         \label{fig:300_nels_all}
   \end{figure}

\begin{figure}
   \centering
   \includegraphics[angle=0,width=\columnwidth]{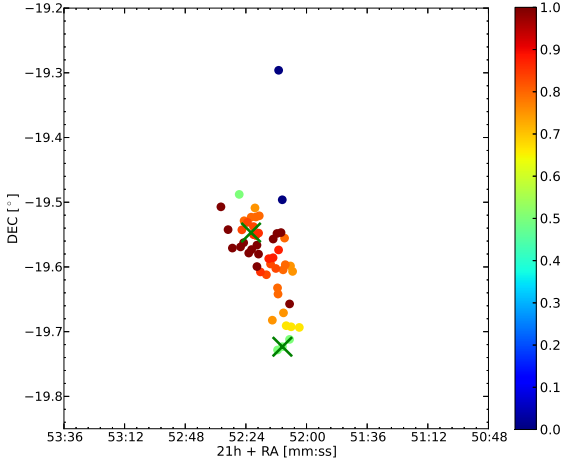}
      \caption{Local fractions of no-EL disk galaxies (with respect to the sample containing all late type galaxies). BCG1 and BCG2 are represented by green crosses. Note the high abundance of no-EL disks in the region between BCG1 and BCG2.}
         \label{fig:300_nels_sp}
   \end{figure}

Among the MCS we find 11 objects identified as late type galaxies (either unambiguously by-eye or confirmed on the basis of their Gini/M$_{20}$ indices) which do not show [OII] or H$\alpha$ emission lines. In the following, we refer to these objects as no-EL disk sample. We also define a peculiar subsample of the MCS which consists of all galaxies showing a disturbed morphology. As shown in Fig. \ref{fig:nelshist} the no-EL disk sample is on average closer to the cluster centre (i.e. closer to BCG1) when compared to the sample of all late type cluster galaxies.\\
Figs. \ref{fig:300_nels_all} and \ref{fig:300_nels_sp} show the local fraction of no-EL disks relative to the MCS and relative to the sample of all late type galaxies, respectively, within a radius of 300 kpc around each galaxy in the MCS. We adopt this visualisation from \citet{pranger13}. The circle size of 300 kpc represents the best compromise between spatial resolution and number of galaxies per resolution element. Note that, for the sake of clarity, the symbol sizes do not correspond to a size of 300 kpc but are much smaller. In compliance with previous studies (e.g. \citealt{vogt04,boesch13}) we find that no-EL disks are concentrated near the cluster centre. They mainly populate the region between both brightest cluster galaxies. Their relative number density is highest in the southern vicinity of BCG1.\\
   
\section{Star formation rates and colours}
\label{sec:sfr}
\begin{figure}
   \centering
   \includegraphics[angle=0,width=\columnwidth]{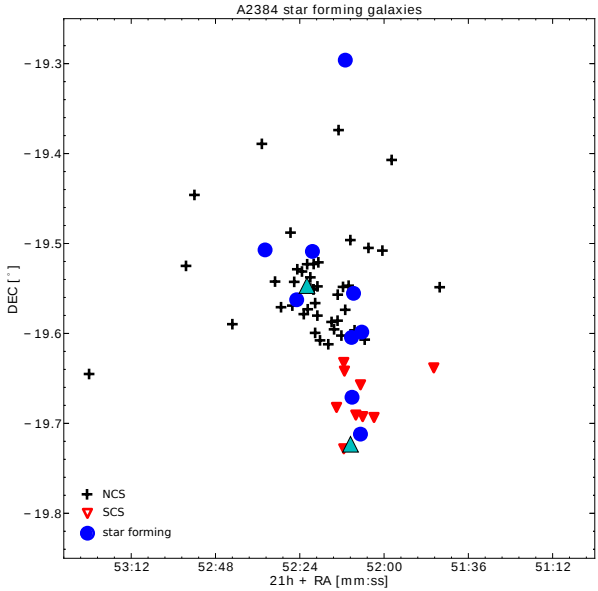}
      \caption{Cluster member galaxies in the inner 30x30 arcminutes of A2384 (centred on BCG1). Black crosses represent galaxies assigned to the NCS, red triangles indicate objects belonging to the SCS. The star forming sample is illustrated by blue circles. Both brightest cluster galaxies (BCG1 and BCG2) are shown as cyan triangles.}
         \label{fig:sfr}
   \end{figure}

\begin{figure}
   \centering
   \includegraphics[angle=0,width=\columnwidth]{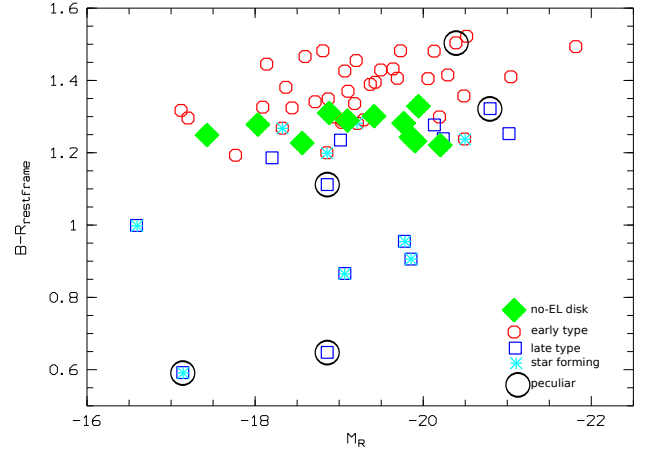}
      \caption{Colour-magnitude diagram of the MCS. Red circles show early type galaxies, blue squares represent late type galaxies and green diamonds illustrate galaxies belonging to the no-EL disk sample (late types). Note that the latter populates the transition region between blue cloud and red sequence. Empty blue squares represent late type galaxies with H$\alpha$ emission line but without detectable [OII] emission line. Superimposed cyan stars mark late type galaxies with [OII] emission line from which we calculate SFRs according to \citet{kennicutt92}. Black circles indicate peculiar objects. The reddest peculiar object is the AGN host and corresponds to galaxy k in the bottom row of Fig. \ref{fig:10_2b}.}
         \label{fig:cmd}
   \end{figure}

\begin{figure}
   \centering
   \includegraphics[angle=270,width=\columnwidth]{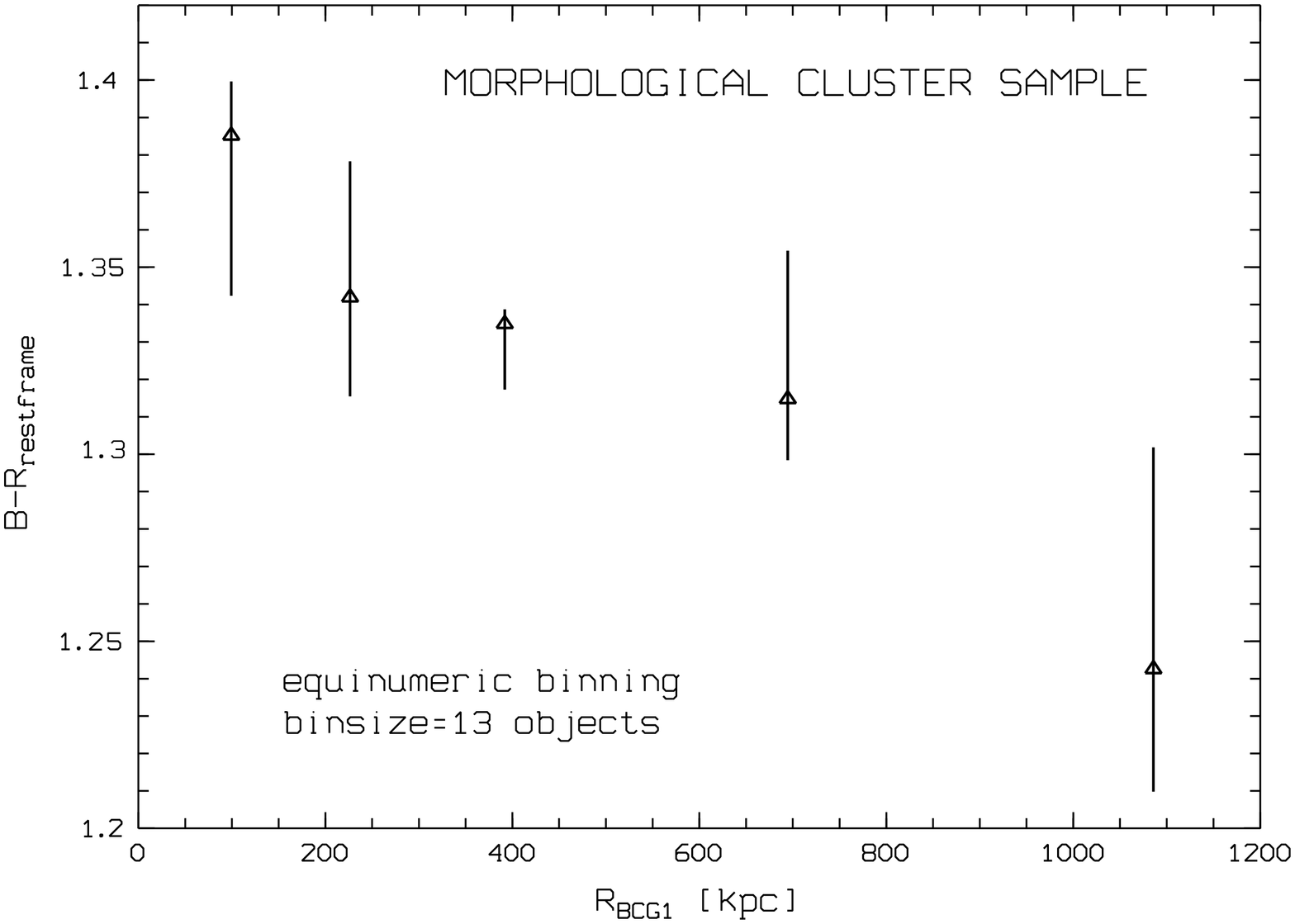}
      \caption{Median B-R restframe colour as a function of clustercentric distance (i.e. distance to BCG1). There is no colour information available on the field sample (which stems from our previous analysis of Abell 3921).}
         \label{fig:col_r}
   \end{figure}

To estimate the star formation rates (SFRs) in the MCS we adopt a relation presented in \citet{kennicutt92}:

\begin{equation}
      $SFR$(M_{\odot}yr^{-1})\simeq 2.7 \cdot 10^{-12} \frac{L_{B}}{L_{B}(\odot)} EW([OII]) E(H\alpha)
      \label{eq:sfr}
\end{equation}

where $EW([OII])$ denotes the rest-frame equivalent width of [OII] and $E(H\alpha)$ stands for the extinction value at the respective wavelength. To apply this measure to our data we first determine the B-band absolute magnitudes $L_{B}$ via apparent magnitudes from the object catalogue generated by \citet{maurogordato11}. In addition, for a proper k-correction all spectra are classified by comparison with spectra from \citet{kennicutt92b}. 10 Galaxies in the MCS have a detected [OII] line. One of the spectra shows strong evidence for an AGN. The corresponding galaxy is not used in our analysis of the SFRs. In the following we refer to the remaining nine galaxies as the star forming sample (SFS). In Fig. \ref{fig:sfr} we show the spatial distribution of both the MCS and, superimposed, the SFS. The latter appears to be evenly distributed over the north-south elongated cluster core. We find star forming galaxies in the northern subcluster (7) as well as in the southern subcluster (2). While eight of the star forming galaxies populate the dense core regions close to the brightest cluster galaxies, the remaining object resides in a low density region at a projected distance of 1.60 Mpc to the north of the cluster centre. In Table \ref{table:tab3} we list SFRs, B-R restframe colour and morphological types for the SFS. The median star formation rate of the SFS is 0.41 $M_{\odot}$/yr which is clearly lower than the field value of $\sim$2.45 $M_{\odot}$/yr.\\ 
In Fig. \ref{fig:cmd} we show the colour-magnitude diagram (CMD) of the MCS split up into late type and early type galaxies. The R-band and B-band aperture magnitudes were calculated in analogy to the B-band absolute magnitudes. We chose an aperture diameter of 2 arcseconds which in our cosmology corresponds to 3.5 kpc at the cluster redshift. In order to rule out potential colour bias we repeated our analysis with an aperture of 3 arcseconds (5.25 kpc) and found only negligible deviations. The no-EL disk sample (which entirely consists of morphological late type objects) is shown separately. Its members populate the transition region between blue cloud and red sequence. We also superimpose peculiar cluster members and the SFS with five star forming galaxies in the late type and four in the early type subsample. Since we calculate SFRs from [OII] equivalent widths we do not have SFR estimates for late type galaxies which show H$\alpha$ line emission but no detectable [OII] emission line (empty squares in Fig. \ref{fig:cmd}). Surprisingly, the galaxy with the highest SFR in our star forming sample is a morphological early type. However, it is the second-bluest early type object in the CMD and it is close to the dividing line in Gini-M$_{20}$ space. In Fig. \ref{fig:col_r} we illustrate median B-R restframe colour as a function of clustercentric distance (i.e. distance to BCG1). The plot was generated in analogy to the radial plots in Sec. \ref{susec:radial}.\\    

\begin{table}    
\centering                          
\begin{tabular}{c c c c}        
\hline\hline                 
SFR & B-R colour & morphological type\\.[$M_{\odot}$/yr]&&\\    
\hline            
\noalign{\smallskip}   
$0.65$ & $1.28$ & early type\\      
\noalign{\smallskip}  
$0.26$ & $1.24$ & early type\\
\noalign{\smallskip}    
$3.30$ & $1.21$ & early type\\
\noalign{\smallskip}   
$0.11$ & $1.26$ & early type\\      
\noalign{\smallskip}  
$0.35$ & $0.87$ & late type\\
\noalign{\smallskip}    
$0.45$ & $0.96$ & late type\\
\noalign{\smallskip}  
$0.38$ & $1.00$ & late type\\      
\noalign{\smallskip}  
$0.44$ & $0.91$ & late type\\
\noalign{\smallskip}    
$0.41$ & $0.59$ & late type, peculiar\\
\noalign{\smallskip}    
\hline                                   
\end{tabular}
\caption{Star formation rates and morphological types of all galaxies in the star forming sample (SFS). 
 }             
\label{table:tab3} 
\end{table}  
  
\section{Discussion}
We determined redshifts of 305 galaxies in the field of the merging galaxy cluster Abell 2384 (z$\simeq$0.094) and combined this sample with redshift data from \citet{maurogordato11} yielding a total sample of 368 redshifts. We defined an interval in redshift space using the common limit of $\pm$5000 km/s relative to the cluster centre in radial velocity and applied a Dressler-Shectman test (DS-test) to all 173 galaxies with redshifts in this interval. The DS-test detected substructure in the east of the field-of-view at a median redshift of $<$z$>$=0.0810. We interpret this structure as a massive group falling onto the cluster from the east (projected onto the plane of the sky) with a high peculiar line-of-sight velocity towards the observer.\\

Applying the EMMIX algorithm on the galaxies in the redshift interval we defined cluster membership and analysed spectroscopical (and photometric) data of $\sim$120 ($\sim$60) cluster galaxies. The results confirm the presence of two brightest cluster galaxies (BCGs) at a projected distance of $\sim$1150 kpc. Both of these giant elliptical galaxies are associated with a dense accumulation of galaxies which we identify as the northern and southern merging subclusters of A2384 (A2384N and A2384S in \citealt{maurogordato11}). We find that these subclusters reside at median redshifts of $<$z$>$=0.0937 and $<$z$>$=0.0944, respectively. Their distributions overlap entirely in redshift space. The median redshift difference of $\Delta$z=0.0007 corresponds to a line-of-sight velocity difference of $\sim$210km/s. A low relative velocity could be the result of dynamical friction acting during the core passage of merging subclusters. Since galaxy-galaxy interactions are most efficient at low relative velocities (see e.g. \citealt{toomre72}) this would be in agreement with the scenario that such interactions cause the occurrence of morphologically distorted galaxies in the cluster core regions. \citet{maurogordato11} found that the most probable collision scenario for A2384 are two clusters seen more than 1.0 Gyr after the first core passage, with a collision direction close to the line-of-sight. Following this hypothesis the 3D relative velocity of the subclusters would be of the order of only a few hundred km/s. Since the mass of the system is still not well determined and the line-of-sight velocity difference of the BCGs is $\sim$1000 km/s (5 times larger than the median difference) this interpretation is somewhat tentative.\\
In a comparison with recent high resolution numerical simulations of galaxy clusters, we performed an analysis of galaxy velocities during and after cluster mergers on the sample of unvirialised cluster-size objects in \citet{hoeller14}. In all such cluster post-merger cases (C06, C10, C11, C12 in \citealt{hoeller14}) we find a number of stable galactic halos with masses greater than $10^{10} M_{\odot}$ and peculiar velocities of only a few hundred km/s. The temporal resolution of the data is of the order of $\sim$40Myrs and analysis shows that the halos decelerated by dynamical friction regain peculiar velocities of $>$500km/s only in a few time steps while the efficiency of this re-acceleration is apparently related to the cluster masses. We find noticeably low relative velocities even in the cluster core regions. Close fly-bys at low relative velocities eventually result into galaxy mergers in our simulations. Before the galaxies actually merge such close low-speed fly-bys will lead to morphological distortions in the individual galaxies due to efficient tidal interactions. Our sample of peculiar galaxies in the core regions of Abell 2384 includes morphologically distorted single galaxies as well as more advanced galaxy mergers. Our simulations are in agreement with these observational findings.\\

Our cluster and subcluster velocity dispersions are in agreement with \citet{maurogordato11} for the total cluster sample and the northern subcluster. For the southern subcluster we obtain a lower velocity dispersion than these authors. This $\sigma$ value is in excellent agreement with the $\sigma - T_{x}$ relation found by \citet{wu98}. Our mass estimates for the total cluster sample (TCS), the northern cluster sample (NCS) and the southern cluster sample (SCS) are probably overestimated due to overestimates in velocity dispersion (for the TCS and the NCS) and mean harmonic radius (for the SCS). The masses we compute exceed the values found by \citet{maurogordato11} by factors of 1.2, 1.3 and 2 for the TCS, NCS and SCS, respectively. Hence we interpret our result for total cluster mass ($2.34^{+0.08}_{-0.07} \cdot 10^{15} M_{\odot}$) as an upper limit. We find an upper limit of $\sim$1.6:1 for the mass ratio of the northern and the southern subcluster. This value is lower than the ratio of $\sim$2.3:1 found by \citet{maurogordato11}. However, this result is still in compliance with their favoured scenario of a merger between a more massive northern and a smaller southern subcluster. As a consequence of probably overestimated velocity dispersions our results for $r_{200}$ of the total A2384 sample and of the northern subcluster are overestimated, too. This is most likely not the case for the southern subcluster ($r_{200}=1761^{+282}_{-499}$ kpc), given its agreement with the $\sigma - T_{x}$ relation.\\

The increasing radial trends in fractions of EL-galaxies and equivalent widths as well as their stagnation after reaching the field level are in compliance with the SF-density relation (see e.g. \citealt{patel11}). However, it is noticeable that for the southern cluster sample the increasing trend only sets in at clustercentric distances $>$3 Mpc and does not reach the field level before $R_{BCG1}\simeq$6 Mpc. This suggests that processes turning star forming galaxies into quiescent galaxies are already efficient at larger clustercentric distances in the southern subcluster.\\

We computed morphological descriptors (Gini coefficient, concentration index, M$_{20}$) and B-R restframe colour for cluster galaxies in the inner 30x30 arcminutes of A2384. For early type galaxies the Gini coefficient as a function of distance to BCG1 is constant at a level slightly higher than the corresponding field value. The concentration index shows an increase between 200 kpc and 400 kpc and stays constant towards larger clustercentric distances. The M$_{20}$ index decreases between 200 kpc and 400 kpc and increases for larger clustercentric distances. While for the concentration index all data points are above the field value we find that all M$_{20}$ data points are below the field value. Standard statistical tests show that none of the weak trends in concentration and M$_{20}$ index is significant.\\
For the late type galaxies we find no significant trends in Gini coefficient and concentration index, albeit the data points show an even larger offset from the field level. Even though we do not find any significant trend within the inner 1.5 Mpc, the offsets from the field levels show that galaxies in the inner cluster regions (and cluster late types in particular) have more concentrated light profiles than their counterparts in the field. The M$_{20}$ values of the late type sample are, however, higher than the field level. We find that this is due to the presence of peculiar galaxies in the morphological cluster sample. After, as a test, removing them from the sample there are only neglectable changes in the positions of the Gini and concentration data points, while the M$_{20}$ data points, within the errors, then comply with the field value.\\
Although our results on spectroscopical and morphological galaxy properties as functions of clustercentric radius are in agreement with earlier findings, we emphasize that our analysis can not distinguish between galaxies which are on their first cluster passage and those which already crossed the core regions (i.e. backsplash galaxies, see e.g. \citealt{pimbblet11}).\\
The presence of peculiar (i.e. morphologically distorted) galaxies in the core region of a merging galaxy cluster is a sign for strong dynamical interactions influencing the shapes of the galaxies' stellar distributions. This hypothesis is in compliance with \citet{kleiner14} who, in their recent analysis of the post merger Abell 1664, find an increase of asymmetry (which results from morphological distortions) in their inner cluster galaxy sample. The authors suggest galaxy-galaxy interactions during the core passage of the smaller merging partner as an explanation for their findings. This interpretation complies with results by \citet{vijayaraghavan13} who, in their cosmological N-body and idealised N-body plus hydrodynamic simulations, showed, that galaxy-galaxy interactions are increased during the core passage of merging subclusters.\\

Within our morphological sample we identify nine star forming cluster galaxies, five of which are classified as late type (including one peculiar galaxy) and four of which are early type objects according to their position in Gini-M$_{20}$ space. \citet{noble13} showed that the inner cluster regions can be contaminated by recently accreted galaxies which are close to the cluster core only in projection and can augment star formation estimates. Regarding the inhomogeneous morphologies of our star forming sample we are, however, motivated to interpret the detection of star formation in the inner cluster region as an imprint of the merging process on the central galaxy population. Due to cluster-merger related dynamical interaction processes (galaxy-galaxy tidal interactions, galaxy mergers) galaxies in the cluster core may show different morphologies. While on a statistical basis star formation is found to be quenched in cluster cores at redshifts z$<$1 such processes might maintain or (re-)induce star formation episodes in certain individual galaxies (see e.g. \citealt{bell06, jogee09, robaina09}). The median star formation of the star forming sample is significantly lower than in the field which is in compliance with the scenario of infalling star forming field galaxies getting quenched by cluster related processes (see e.g. \citealt{vonderlinden10}).\\        

Similar to other investigations of low - to intermediate - redshift galaxy clusters (e.g. \citealt{boesch13, pranger13}) we find a population of galaxies morphologically identified as late type but not showing any star formation (no-EL disks in our nomenclature) in the cluster core region. These objects are on average closer to the cluster centre (i.e. closer to BCG1) than the sample of all late type cluster galaxies in A2384. Also other authors (e.g. \citealt{goto03, koopmann04}) found no-EL disks (also referred to as "passive" or "quenched" disks) mainly in high-density environments. In accordance with the generally adopted scenario, we propose a cluster related mechanism, namely ram-pressure stripping, to explain our findings. No-EL disks close to the cluster centre have been subject to ram-pressure stripping during their cluster infall. Hydrodynamic simulations by \citet{quilis00} showed that ram-pressure stripping can remove the majority of a galaxy's atomic hydrogen content on a time scale of $\sim$100 Myr. As a result, star formation will be quenched. In their analyses of 329 nearby cluster and field spiral galaxies \citet{vogt04} proposed no-EL disks (or "quenched" disks) as an intermediate stage of a morphological transformation process turning infalling field spirals into S0 cluster galaxies. Although the resolution of our imaging data is insufficient to identify S0 galaxies, the occurrence of no-EL disks close to the cluster centre is in agreement with the transformation scenario.\\ 

Plotting B-R restframe colour of the cluster galaxies in the inner 30x30 arcminutes of A2384 against clustercentric distance we find a negative trend, i.e. a colour-density relation (see e.g. \citealt{cooper07}). Our colour analysis has further shown that the sample of no-EL disks populates the transition region between blue cloud and red sequence (green valley) in the colour-magnitude diagram. In their investigations of red spirals in the galaxy cluster system Abell 901/902 \citet{wolf09} show that, on average, red spirals have a specific star formation rate four times lower than that in blue spirals. The authors also find that their distribution of red spirals in the inner cluster regions populates the green valley in colour-magnitude space. In a more recent study on the quenching of star formation in low-redshift galaxies \citet{schawinski14} came to the conclusion that in principle, galaxies can move through the green valley in both directions (i.e. towards blue cloud \textit{and} towards red sequence). Furthermore, the path along which late type galaxies get quenched does not necessarily have to end in a truncation or destruction of the stellar disk. This is in compliance with the occurrence of galaxies with undisturbed stellar disks in the green valley in colour-magnitude space. Moreover, \citet{schawinski14} state that \\ 
In a follow-up kinematic analysis in Abell 901/902 \citet{boesch13b} found that red spirals have particularly high rotation-curve asymmetries, suggesting an enhanced effect of ram-pressure. We argue that no-EL disks probably are the successors of red spirals. At a later stage of the transformation sequence these no-EL disks might be turning into S0 galaxies.\\

\citet{maurogordato11} suggest that Abell 2384 might be a post-merger system where two cool-cores have survived the first core passage. The authors refer to \citet{poole06} who find in their numerical simulations that initial cool-cores of merging clusters can survive the first pericentric passage but disappear after the second crossing.\\
In \citet{hoeller14} we conducted high resolution ICM simulations with realistic metal enrichment evolution utilizing a sophisticated subgrid model for galactic winds and ram pressure stripping based on the semi-analytical galaxy formation model \textsc{Galacticus} \citep{benson12}. We find that the formation of cool-cores strongly depends on the ICM metallicity and its evolution. In a comparison of three galactic wind prescriptions within these cosmological simulations, we have shown that the model which fits observational data best in terms of star formation histories and ICM metallicities, strongly favours the formation of cool-cores. Moreover, these cool-cores are not necessarily disrupted by merger events if they form at a sufficiently early stage of cluster formation. Cluster C01 in \citet[see Fig. A.1.(a), 
A.2.(a) and A.3.(a)]{hoeller14} has experienced an off-center merger event at z$\sim0.3$ and shows two elongated cool-cores associated with two BCGs and their relative trajectories. The very energetic merger in cluster C10 has completed a second central passage at z$=0$ and shows an extended central cool plateau embedded in very hot, shock-heated ICM \citep[see Fig. A.2.(j)]{hoeller14}. The best candidate for a merger-induced temperature bimodality as seen in the temperature maps of Abell 2384, is cluster C11. At z$=0$ two arcs are emerging from the center perpendicular to the merger direction, one carrying cool gas while the other contains comparably hot gas \citep[see Fig. A.2.(k)]{hoeller14}. In view of 
the numerical data, we can support the conclusions drawn in \citet{maurogordato11} concerning the dynamical history of Abell 2384.

\subsection{Comparison with Abell 3921}
In this section we will compare the results presented in this paper with our analysis of the galaxy population in the merging cluster system Abell 3921 \citep{pranger13}. Both clusters are bimodal merging systems at a redshift of z$\simeq$0.094.\\ 

\begin{figure}
   \centering
   \includegraphics[angle=270,width=\columnwidth]{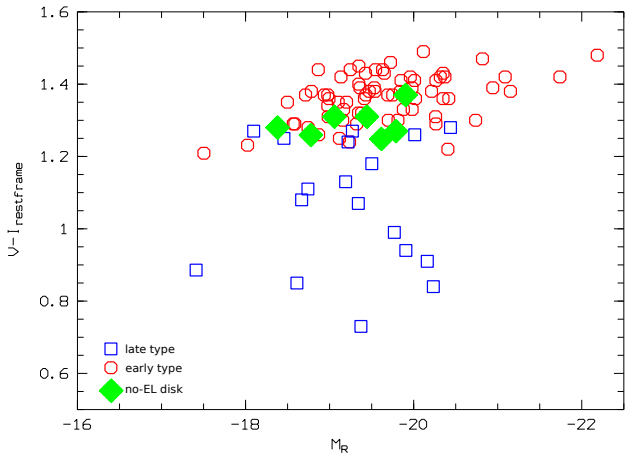}
      \caption{Colour-magnitude diagram of Abell 3921. Plotted are galaxies populating the inner $\sim$3x3 Mpc$^{2}$ (same as in Fig. \ref{fig:cmd}). Red circles illustrate early type galaxies, blue squares show late type galaxies and green diamonds represent no-EL disks.}
         \label{fig:comp}
   \end{figure}

In both sky areas we find substructure which might be massive groups falling onto the cluster with a high peculiar line-of-sight velocity. Moreover, in both systems we find that spectroscopic galaxy parameters (fraction of galaxies showing [OII] or H$\alpha$ emission lines, [OII] and H$\alpha$ equivalent widths) in general increase with clustercentric distance, regardless of the ongoing cluster merger. Both clusters host populations of no-EL disks close to their centres. The fraction of no-EL disks with respect to all galaxies in the cluster core regions are 7,4$\%$ and 17,7$\%$ for A3921 and A2384, respectively. In both clusters the sample of no-EL disks is on average closer to the cluster centre than the sample of all late type galaxies. Both no-EL disk samples populate the transition region between blue cloud and red sequence in the colour-magnitude diagram (see Figs. \ref{fig:cmd} and \ref{fig:comp}).\\

While in Abell 2384 the trends in the spectroscopic parameters as functions of clustercentric distance are monotonic, we find local decreases at $\sim$3.5 Mpc clustercentric radius in Abell 3921. This is due to the occurrence of no-EL disks in these regions. \citet{pimbblet06} present similar results on Abell 3921 and suggest pre-processing in infalling substructurs as an explanation for this occurrence. In \citet{pranger13} we argue that no-EL disks at large clustercentric radii could at least partially be explained by merger shock waves in the intra cluster medium (ICM). When these shock waves move through a galaxy (or vice versa) they induce increased ram-pressure which might at first lead to an increase and on longer timescales a quenching of star formation activity (see eg. \citealt{kapferer09, quilis00}). In their analysis of two merging galaxy clusters hosting radio relics which trace ICM shock waves, \citet{stroe14} find strong signs for shock-induced star formation in galaxies close to the radio relic in the less advanced cluster merger with $t_{0}\lesssim$1.0 Gyr ($t_{0}$=0 denotes the time of coalescence). Recent simulations of the shock wave scenario by \citet{roediger14} show that star formation is indeed enhanced by shock waves, albeit only at galactocentric radii where the gas will be stripped in due course. These results are both in agreement with our line of argument presented in \citet{pranger13}. Since A3921 is most probably a pre-merger with $t_{0}\simeq$-0.3 Gyr and A2384 a post-merger with $t_{0}\gtrsim$1.0 Gyr \citep{maurogordato11, kapferer06}, the time-separation in dynamical evolution between both clusters is $\gtrsim$1.3 Gyr. If we interpret the current dynamical states of A3921 and A2384 as "snapshots" of comparable merging scenarios, we can conclude that the higher fraction of no-EL disks found in the core regions of A2384 (w.r.t. A3921) is a consequence of a more advanced merger state. In A2384, more infalling spiral galaxies, including the ones which might have been ram-pressure-quenched (i.e. turned into no-EL disks) by shock waves at larger clustercentric distances, have had time to populate the cluster core regions. Note that we could not investigate galaxy morphologies at clustercentric distances greater than 1.7 Mpc in A3284 because of the limited coverage of our imaging. However, we do not find no-EL disks in A2384 at clustercentric radii greater than 880 kpc. To explain this we argue that at the advanced dynamical state of the merger in A2384, merger shock waves might already have fainted and/or moved further away from the inner cluster regions. This interpretation is in compliance with \citet{stroe14} who do not find any traces of enhanced star formation activity near to the radio relic in the older of the two mergers they analyse ($t_{0}\simeq$2.0 Gyr).\\    
While in Abell 3921 we do not find any morphologically distorted (i.e. peculiar) galaxies in the inner cluster region ($\sim$3x3 Mpc$^{2}$), $\sim$8$\%$ of the galaxies in the inner cluster region of Abell 2384 are peculiar objects. Since Abell 3921 is in a pre-merger phase while Abell 2384 is already in a post-merger state, this complies with the hypothesis that cluster mergers can increase the probability for galaxy-galaxy interactions in the cluster core.\\

\section{Conclusions}
   
In summary, our spectrophotometric analyses of Abell 2384 motivate the following conclusions:

\begin{itemize}
\item
The substructure detected in the east of the central cluster region might be a massive group falling onto the cluster from the east (projected onto the plane of the sky) with a high peculiar line-of-sight velocity towards the observer. This is motivated by an offset of $\Delta$z=-0.013 with respect to the systemic cluster redshift and a group-characteristic velocity dispersion of $\sim$350 km/s.  
\item
The radial trends of the spectroscopic quantities and B-R colour are in agreement with the SF-density relation and the colour-density relation, respectively. The onset of the trends in the southern subcluster only at distances $\geq$3 Mpc suggests that cluster specific quenching processes (e.g. ram-pressure stripping) are efficient already at large clustercentric radii.  
\item
The occurrence of peculiar (i.e. morphologically distorted) galaxies in the cluster core regions is in compliance with the post merger hypothesis according to which peculiar galaxies close to the cluster centre are a consequence of frequent tidal galaxy-galaxy interactions and galaxy mergers that happened during the core passage of the southern subcluster. This hypothesis is supported by our numerical simulations.
\item
The morphological mix of the star forming galaxies in the inner $\sim$3x3 Mpc$^{2}$ of Abell 2384 can be interpreted as an imprint of the cluster merger. Galaxy-galaxy interaction processes may have maintained or (re-)induced star formation episodes in galaxies of different morphological type.
\item
We detect a population of no-EL disks close to the cluster centre. This is in agreement with the scenario of morphological transformation in regions of increased density. It suggests that no-EL disks represent an intermediate stage in the transition of infalling field spirals into cluster S0s. This interpretation is strengthened by the position of our no-EL disk sample in the colour-magnitude diagram. 
\end{itemize}

\begin{acknowledgements}
We thank the anonymous referee for the constructive comments which helped a lot to improve the manuscript. 
We are very grateful to Scott Croom and Ivan Baldry for having provided the codes \texttt{runz} and \texttt{autoz}. Florian Pranger, Asmus B\"{o}hm and Sabine Schindler are grateful for funding by the Austrian Funding Organisation FWF through grant P23946-N16. Chiara Ferrari acknowledges financial support by the "\textit{Agence Nationale de la Recherche}" through grant ANR-09-JCJC-0001-01. 
\end{acknowledgements}    

\bibliographystyle{aa}
\bibliography{flo_bib}

\end{document}